\documentclass[12pt,preprint]{aastex}

\slugcomment{Resubmitted to ApJ  6/7 Dec 2005}

\shorttitle{ACRIMSAT Transit of Venus}
\shortauthors{Schneider, et al.}

\begin{document}

\title{The Effect of the Transit of Venus on ACRIM's Total Solar Irradiance Measurements: Implications for Transit Studies of Extrasolar Planets}

\author{G. Schneider}
\affil{Steward Observatory, The University of Arizona, 
Tucson, AZ 85721}
\email{gschneider@as.arizona.edu}

\author{J. M. Pasachoff}
\affil{Williams College, Williamstown, MA 01267}
\email{jay.m.pasachoff@williams.edu}
\and

\author{Richard C. Willson}
\affil{Columbia University, New York, NY 10027}
\email{rwillson@acrim.com}

\begin{abstract}
We have used the 8 June 2004 transit of Venus (ToV) as a surrogate to test observing methods, strategies and techniques that are being contemplated for future space missions to detect and characterize extrasolar terrestrial planets (ETPs) as they transit their host stars, notably NASA's {\it Kepler} mission planned for 2008.  As an analog to ``{\it Kepler}-like'' photometric transit observations, we obtained (spatially unresolved) radiometric observations with the ACRIM 3 instrument on ACRIMSAT at a sampling cadence of 131 s to follow the effect of the ToV on the total solar irradiance (TSI). Contemporaneous high-resolution broadband imagery with NASA's TRACE spacecraft provided, directly, measures of the stellar (solar) astrophysical noise that can intrinsically limit  such transit observations.  During the Venus transit, which  lasted $\sim$ 5.5 hours, the planet's angular diameter was approximately 1/32 the solar diameter, thus covering $\sim$ 0.1\% of the stellar surface. With our ACRIM 3 data, we measure temporal changes in TSI with a 1 $\sigma$ per sample (unbinned) uncertainty of approximately 100 mW m$^{-2}$  (0.007\%). A diminution in TSI of $\sim$ 1.4 W m$^{-2}$ ($\sim$ 0.1\%, closely corresponding to the geometrically occulted area of the photosphere) was measured at mid-transit compared with a mean pre/post transit TSI of $\sim$ 1365.9 W m$^{-2}$. The radiometric  light curve is complex because of the parallactic motion of Venus induced by ACRIMSAT's near-polar orbit, but exhibits the characteristic signature of photospheric limb-darkening. These observations serve as a surrogate to future photometric observations of ETPs such as {\it Kepler} will deliver. Detailed analysis of the ToV, a rare event within our own solar system, with time-resolved radiometry augmented with high-resolution imagery 
provides a useful analogue for investigating the detectability and characterization of ETPs from observations that are anticipated in the near future. 

\end{abstract}


\section{INTRODUCTION}

Since the seminal detection and identification of the companion to 51 Peg as a giant planet orbiting very close to its host star \citep{Mayor95}, the past decade has seen an explosion of detections of extra-solar ``hot Jupiter'' type planets around solar-like stars\footnote{http://vo.obspm.fr/exoplanetes/encyclo/catalog-main.php; http://exoplanets.org/}.  Direct detection of young ($<$ few tens of Myr) gravitationally-bound companions of planetary mass (predicated on evolutionary cooling models) to nearby ($<$ few tens of parsecs) stars via high-contrast space-based coronagraphy (e.g., \cite{Schn03}) and/or ground-based adaptive optics augmented imaging in the near-IR  (e.g., \cite{Mac01}) is now technically feasible.  Detection (and eventually characterization)  of terrestrial planets around solar-like stars at an advanced stage of evolution and  in their ``habitable zones'' remains a challenging, though yet unfulfilled, key goal in observational astronomy and astrophysics. Major space-based initiatives such as the {\it Kepler} \citep{Bor03}, and even more technically challenging {\it TPF} \citep{Bei98}, missions are under way and development will, one hopes,  will rise to this challenge. 

Indirect detection of Jovian-mass extrasolar planets has been demonstrated by the transit method (e.g.: HD 209458b, \cite{Char00}; TrES-1, \cite{Alonso04}). Fledgling steps have been taken toward exoplanet atmospheric identification and characterization by  transit spectroscopy as well (e.g., HD 209458b with the detection of atmospheric sodium, \cite{Char02}) and conceptually may be extended, with a much higher degree of observational complexity, to ETPs yet to be found. The recent predicted transit of an  known terrestrial planet in the habitable zone of a {\it very} nearby solar-``like'' star,  i.e. the transit of  Venus across the Sun on 8 June 2004 \citep{Pas05}, provided a unique and timely opportunity to validate the concepts, instrumentation, observational processes and procedures that will be developed over the next decade to extend our vision of terrestrial planets to other solar systems. Seizing on that rare opportunity we obtained space-based solar irradiance (total irradiance power) measurements  with the Active Cavity Radiometer Irradiance Monitor\footnote{http://www.acrim.com/} (ACRIM) 3 instrument on ACRIMSAT to evaluate the radiometric (closely analogous to broad band photometric) detectability of spatially unresolved ETP transits of their host stars using the ToV as a ``nearby'' analog.  With contemporaneously obtained high-resolution broadband imagery of the transit obtained with the Transition Region and Coronal Explorer\footnote{http://trace.lmsal.com/} (TRACE) spacecraft to evaluate the intrinsic stellar (i.e., Solar) ``astrophysical'' noise, we compare transit light-curve derived systemic results with our  {\it a priori} knowledge of the characteristics of the Venus/Sun system.

\section {OBSERVATIONS and INSTRUMENTATION}

Both the ACRIMSAT and TRACE spacecraft provide unique ``views'' of the Sun from low Earth orbit (LEO), each offering instruments well-suited to this investigation.  The ACRIM 3 radiometer \citep{will} provides the total (0.2 to 2.0~$\mu$m) solar irradiance (TSI; measured in W m$^{-2}$ received at  1 AU).  The White Light (WL) channel of the TRACE imager \citep{strong} provides high ($\sim$ 0\farcs5) spatial resolution broadband (0.1 to 1.0 ~$\mu$m) solar photospheric imagery. While TRACE observes the Sun continuously, ACRIMSAT's observations are interrupted for $\sim$ 30 m by Earth occultation during each of its $\sim$ 100 m orbits.

\subsection {ACRIM 3 Radiometry}

ACRIM 3 was designed to provide accurate, highly precise, and traceable radiometry over decadal timescales, to detect changes in the total energy received from the Sun by the Earth (e.g., Fig. 1). ACRIM 3's  long-term radiometric traceability (including self-calibration of sensor degradation) is 0.0003\% yr$^{-1}$.

The basic TSI data obtained by ACRIM 3 are the average of 32 s of sampling (at a 1.024 s cadence) during each set of shutter open (observations) and closed (calibration) measurements. While standard data products are binned into daily means, ACRIM 3 also provides shorter timescale  calibrated measures (a ``shutter cycle'' read out) with a 1~$\sigma$ single-sample uncertainty of $\sim$ 0.1\% and a precision of $\sim$ 0.01\% (100 micromagnitudes in the extremely broad radiometric ACRIM 3 ``pass band'').  The ACRIM 3 instrument delivers these measures  once every 131.072 seconds.  The radiometric data we discuss in this paper were derived from these measures which were obtained during, and flanking, the 8 June 2004 ToV from 01h UT to 16h UT. 

\subsection{TRACE Imagery}
Using the TRACE WL channel, which provides spectral sensitivity in the wavelength range from 0.12 - 0.96 $\mu$m, we obtained 98 image frames during the transit ingress from 050553 UT to 054751 UT and 84 frames during egress from 105231 UT to 112315 UT. Because of the need for high inter-frame temporal cadence in the highest fidelity available data format, we read out only a 512 x 512 sub-array of the 1024 x 1024 CCD detector due to both downlink bandwidth  and memory (image buffer) limitations.  Our observations were interleaved with those of other programs (using other TRACE filters and data formats), thus the inter-frame temporal spacing was not uniform throughout our imaging sequences. During ingress we achieved a cadence of five frames per minute (uniformly spaced) across the limb contacts, but at a reduced rate of $\sim$ two frames per minute at other times. At egress we obtained images at third contact $\pm$ 3 m at a uniformly spaced rate of 7 frames per minute, but at a slower rate of $\sim$ one frame every 35 s when the leading edge of Venus was further from the solar limb.  The image data were processed as described by \cite{Schn04}, where additional information on the TRACE WL channel for planetary transit imaging may be found.  

The TRACE WL optical channel, so we discovered, suffers from an $\sim$ 1\% intensity optical ``ghosting.''  Normally, this is not of concern for TRACE's solar observation programs, but  our interest rested, primary, in the bottom 1\% of the (12 bit) dynamic sampling range.  With detailed characterization (including the analysis of identically acquired pre-transit imagery) we found the ghosting to be double, with different degrees of afocality and intensity, and an additional diffuse (not afocally specular) component. For each processed image frame, we built and subtracted models of the instrumentally scattered light to remove these artifacts before photometric measures were later made\footnote{A full description of the ghost removal process, along with representative post-processed image frames after additionally removing the radial intensity gradient induced by solar limb-darkening, may be found at: http://nicmosis.as.arizona.edu:8000/ECLIPSE\_WEB/TRANSIT\_04/TRACE/TOV\_TRACE.html}. 

During the transit, Venus's angular diameter was 58\farcs2 (12,104 km at 0.289 AU), diametrically spanning 116 TRACE detector pixels.  The TRACE WL spatial resolution is Nyquist-limited by its 0\farcs5 pixel$^{-1}$ sampling of its CCD, not diffraction limited by the 0.3 m diameter telescope.  With a WL flux-weighted mean wavelength of 0.62 $\mu$m, Venus was resolved by $\sim$ 60 band-integrated resolution elements.

\section{The ACRIMSAT ORBIT}
ACRIMSAT is in a Sun-synchronous orbit, but, unlike TRACE, it cannot observe the Sun continuously. ACRIMSAT's line-of-site to the Sun is occulted by the Earth during each spacecraft orbit.  As a result of its orbital geometry ACRIM 3 is actively measuring the TSI for about an hour orbit in each of its approximately 100 minute orbits.  These periodic interruptions are not of concern to the primary mission for ACRIM -- to monitor with higher precision long-term variations in the solar radiometric output.  This does mean, however, our ToV data set which otherwise nominally provides samples once every 131.072 seconds, ``suffers'' from intra-orbit gaps due to visibility interruptions.

The line-of-site solar visibility intervals and Venus transit circumstances, as seen by ACRIMSAT, were determined for time intervals spanning the ToV using a definitive (post-priori) orbital ephemeris derived from a contemporaneous epochal satellite element set provided by NORAD.

\subsection{ACRIMSAT's ``View'' of the Transit}
The apparent path of (the center of) Venus as seen from ACRIMSAT with respect to the heliocenter as the planet crossed the face of the Sun is depicted in Figure 2.  The planetary parallax (shifting of the line-of-sight to Venus) induced by the spacecraft orbit, projected onto the disk of the Sun, causes periodic spatial and temporal modulations in the location of Venus as it traverses the solar disk.  The ``vertical'' amplitude variations (i.e., in the North/South direction in this ``North up'' illustration)  result from the near-polar ACRIMSAT orbit. As Venus is of similar size as the Earth, and as ACRIMSAT is in a low Earth orbit, the vertical excursions are also comparable to (but a bit larger than) the diameter of Venus.  The ``horizontal'' (East/West) component manifests itself in non-linear spacings in the planetary position along its projected path in equal time intervals.  This variation results from the ACRIMSAT orbit plane not being in the line-of-sight direction to the Sun. With TRACE the modulation is more closely sinusoidal, as its orbit plane is perpendicular to the Earth/Sun line, but deviates from a true  sinusoid (linear in spacecraft orbital phase angle with time) because of the planet's orbital motion about the Sun.

\section{The ACRIM 3 Radiometric ``Light Curve''}
The disk of Venus, fully silhouetted in front of the Sun as viewed by ACRIMSAT during the ToV, occulted $\sim$0.1\% of the total area of the photosphere, so a radiometric detection of the transit by ACRIM 3 was readily expected -- and was obtained. The ACRIM 3 ToV data, provided and radiometrically calibrated by the ACRIM Experiments team (extending  5 h on the external sides of the transit ingress and egress), is shown in Figure 3 as radiometric light curve. Gaps in the ACRIM 3 light curve primarily result from Earth occultations, though a small amount of data was irreparably lost elsewhere in the downlink path from the spacecraft. Unfortunately, the transit egress (the interval between Contacts III and IV) was not visible from ACRIMSAT. 

TSI observations were also obtained from the Total Irradiance Monitor (TIM) instrument on the Solar Radiation and Climate Experiment (SORCE) spacecraft\footnote{http://spot.colorado.edu/$\sim$koppg/TSI/}$^,$\footnote{http://www.gsfc.nasa,gov/news-release/releases/2004/h04-235.htm}. Coincidentally (and unfortunately), the phasing of the SORCE orbit w.r.t. its periods of Earth occultation during the ToV was very similar to ACRIMSAT's and egress was also unobserved by TIM. As we use only the ACRIM 3 data in our analysis we do not discuss the TIM data further.  We note, also, that the SOHO spacecraft (in a halo orbit about the L1 Lagrangian point rather than LEO) was not in a zone from which the transit was visible, thus transit observations from the VIRGO instrument were not possible.

\section{A MODEL LIGHT CURVE}
We tested the recoverability of the {\it a priori} known transit parameters (i.e., the geometrically derived transit depth and contact times, knowing d$_{Venus}$/d$_{Sun}$ and positions of Venus with respect to the heliocenter as a functions of time given the spacecraft and planetary orbits). To do so we constructed a model light curve by building a series of two-dimensional synthetic transit images (e.g., see Figure 4).  A circular sub-aperture (representing the disk of Venus) geometrically ``occulted''  portions of the model solar disk as the planet transited the photosphere. The model solar image was built, parametrically representing the center-normalized radially limb-darkened surface profile, SB(r), with a simple two-parameter form:
  SB(r) = 1 -  {\it a}(1-[sqrt(1-r$^2$)]$^{b}$);   where r is the fractional solar radius,
as suggested by \cite {Hestroffer}.

\subsection{Light Curve Asymmetries}
As a result of the reflective spacecraft parallactic motion of Venus, the planet's apparent path across the Sun ``nods'' in heliocentric radius (r) as illustrated in Fig 2.  This motion induces  asymmetries in the ACRIM radiometric light curve as Venus occults portions of the solar disk of differing surface brightnesses (flux densities) in a radially dependent manner due to solar limb darkening.  One can see that during ingress Venus crosses from r = 1.0 to r = 0.9 (where the limb darkening function has a very steep gradient, see Figure 5) twice as slowly as it does from r = 0.9 to r = 1.0 upon egress.  Hence, the downward slope of the ingress light curve is more shallow than during egress.  Additionally, small amplitude, orbit periodic variations in the TSI  measured by ACRIM are expected as Venus oscillates between the brighter (smaller r) portion of the photosphere and positions closer to the solar limb (larger r). This is, at least in part, the cause for some of the ``wiggles'' which are seen at the bottom of the light curve.  Aperiodic variations may result from intrinsic changes in the TSI over the same time interval, or may arise as Venus occults isolated regions of the photosphere differing in local surface brightness (e.g. sunspots or smaller spatial scale localized features). Inspection of contemporaneous high-resolution broadband imagery from the TRACE spacecraft suggests that both effects are seen during the course of the transit.  We will later consider their statistical significance in the ACRIM 3 data set, and how well the orbit periodic effect is separable from intrinsic variations in the global TSI over the same time interval, and also as Venus occults photospheric regions that may be intrinsically brighter or dimmer, in the context of TRACE imagery.

\subsection{Limb Darkening \& Spacecraft Orbit Parallax}
A statistically significant shallow diminution in the radiometric flux density is seen after second contact but before mid transit, i.e., approximately -0.04\% at 05:50 UT and approximately -0.08\% at 06:15 UT compared to approximately -0.10\% at mid transit, and a corresponding gradual rise before the loss of data due to Earth occultation upon egress.  This effect is fully attributable to radially differentiated solar limb darkening, with a strong photospheric radial surface brightness gradient as the limb of the Sun is approached. At 05:50 UT the center of Venus was $\sim$ 0.933 solar radii from the heliocenter, whereas at mid-transit ($\sim$ 08:35 UT) the center of Venus was 0.650 solar radii from the heliocenter. This non-linearity in impact distance with time arises, primarily, from the modulation in Venus's heliocentric velocity vector w.r.t. the limb (i.e., affecting the ``limb crossing angles'') induced by ACRIMSATs orbital parallax.

\section{LIGHT CURVE FITTING}

Model light curves were fit to the observed TSI data via iterative, damped, non-linear least-squares differential corrections (e.g., \cite{Schn85}), analogous to the method used by \cite {Schultz} in solving for the systemic parameters of the HD209458A/B system using {\it Hubble Space Telescope}/Fine Guidance Sensor transit light curves (with some differences in detail noted in \S 5). Geometrically, however, extrasolar planetary systems (such as HD209458A/B) are effectively ``at infinity'' as seen from the Earth, so the ratio of planet:star angular and physical diameters are identical. This is not the case for the Venus transit geometry because of the close proximity of both objects to the Earth.  Because of this, and the added non-linearity in transit geometry due to the spacecraft-orbit induced planetary parallax, rather than adopting a parametric representation such as discussed by \cite{Schultz} (which includes the time of mid-transit, ratio of planet to stellar radii, and orbital inclination), we fit the geometrical times of transit contacts (CI -- CIV),
the  transit depth (time-dependent decrement in TSI), and the solar limb darkening coefficients {\it a} and {\it b} as free parameters (each dependent upon impact distance r/r$_{sun}$).  The time differentials CII-CI (duration of ingress) and CIII-CIV (duration of egress) were not constrained to be equal, as asymmetry in the observed light curve was expected due to the ACRIMSAT orbit-periodic parallax effects previously discussed.  The light curve model parameters which best fit the observations along with their formal 1~$\sigma$ errors (after iterative convergence), are given in Table 1 (column labeled ``measured'').

\subsection{Transit Timing and Error Estimation}
Unlike a true extrasolar planetary transit, the measured times of contacts, and thus ingress, egress, and transit durations, are readily verifiable from our {\it a priori} knowledge of the Venus's (and ACRIMSAT's) orbits.  The derived UTs of the egress contacts from our light-curve fitting, particularly CIV, are less well determined than for ingress. This is primarily  due to the data loss from Earth occultation during CIII--CIV, but also because Venus's angular velocity component in the direction perpendicular to the solar limb was larger at egress than ingress (see Fig 2).  None-the-less, in all cases the  contact times determined from the TSI measures are consistent (within their with 1~$\sigma$ fitting uncertainties) with the back-predicted UTs from the definitive ACRIMSAT orbital ephemeris.

The measured durations of ingress (CII-CI) and egress (CIV-CIII) are both in error (but in opposite sense) by about 23.5\% from the true durations, which correspond to equivalent fractional errors in estimations of the planetary radius. If averaged this would reduce to a near zero error, but this is a coincidence of small sample statistics.  A simple estimation of the expected uncertainty, in the absence of (in this case non-existing) multi-epoch observations, is predicated on  combining the measurement uncertainties in quadrature, i.e. $\pm$ 22.0\% (1~$\sigma$). This is in very good agreement with the percentage errors found in comparing the observationally determined, and ephemeris-predicted, contact times.

The full duration of the transit, measured as the time differential between the mid-points of ingress and egress, is formally uncertain by 2.4\%. This  corresponds to an expectation of uncertainty in the determination of the ratio of the planetary to stellar radii.  The much smaller percentage error in the transit duration determination from the TSI light curve fitting is, again,  just a happenstance of the nearly compensating measurement errors at ingress and egress.

From these data we would expect similar uncertainties in photometricaly derived systemic parameters from extrasolar planetary transits - if sampled equivalently and measured with equal precision.  Improvements would be expected with multi-epoch (repeated) observations and with higher S/N measurements for equivalent per-sample integration times. In the absence of instrumental systematics, stellar variability and non-radial photospheric surface brightness (PSB) variations may ultimately limit transit detection thresholds for ETPs and the uncertainties in derived systemic parameters for ETPs above those thresholds.

\subsection{Intrinsic (Astrophysical) ``Noise''}

We used the as-measured TSI variations in the flanking out-of-transit radiometry to assess ``how good'' (or deficient) our  model light curve fits the data, i.e., how much of the fit residuals are due to instrumental measurement errors and intrinsic solar variations compared to imperfections in the model itself.  ``Variations'', here, not only include temporal variations in (area integrated) TSI but also spatially as Venus covers different parts of the photosphere that are not isotropic in intensity on small spatial scales.  

In Figure 6 we show the difference in the dispersion in in-transit compared to pre/post-transit model light curve fit residuals, which is +25\% in ``as measured'' TSI variability (after subtracting out the model light curve).  This in-transit increase in the dispersion in TSI of 0.002\% during the transit is an order of magnitude larger than the dispersions about the median as-measured TSI's before and after the transit. One may posit one or more instrumental (1), systematic (2), and or real physical effects (3 and 4) contributing to this increase as delineated below:  

(1) Uncertainties in the end-to-end wavelength-dependent system responsivity function for ACRIM under its very broad pass band  (i.e., its spectral sensitivity).  These uncertainties are likely insignificant based upon the ACRIM 3 cavity design and pre-launch testing.

(2) Insufficient fidelity in the limb darkening model.  A quadratic model may be better, and could be tested, but a higher order (multi-parametric) model is likely unjustified given the interrupted phase coverage and single-epoch-only nature of the light curve.

(3) The effects of the atmosphere of Venus itself (absorption, aerosol scattering and refraction).

(4) The effect of Venus occulting regions of the photosphere differing in brightness on small spatial scales.

Neglecting or better characterizing (1), a higher order limb-darkening model (2) could be considered and tested by parametric variation bounded by the instrumental spectral sensitivity calibration. With that, rigorous detection limits for the planetary atmosphere (3) might be ascertained, within the uncertainties in the local variations in photospheric surface brightness on spatial scales of the diameter of Venus (4).

\section{A SURROGATE TO AN ETP TRANSIT}
The solar photosphere is non-isotropic in its surface brightness distribution on spatial scales both smaller, and larger, than the apparent diameter of a Venus. Hence, variations in measured TSI will arise during the transit that depend upon the apparent planet/solar crossing geometry. ACRIM 3, however, sees the Sun as a spatially unresolved source, so spatial (transit geometry dependent) and temporal variations in TSI cannot be decoupled from the radiometric data alone. This will also be the case for photometric transit light curves of ETPs (such as those to be obtained by the {\it Kepler} mission). In the ÒtestÓ case of the transit of Venus, as a surrogate to an ETP transit, that degeneracy can be broken with contemporaneous high-resolution imaging of the solar photosphere.

At the time of the transit, between contacts II and III, the planetary disk of Venus occulted 0.0942\% of the solar photosphere. But, with an optically thick atmosphere extending to $\sim$ 60 km  above the Cytherian surface (very high opacity up to the mesospheric cloud layer), the areal coverage was 0.0961\%. Thus, Venus's atmosphere effectively blocked an additional 0.002\% of the received TSI (if not preferentially forward scattered, refracted, or re-radiated by the atmosphere).  We tested the ability to discriminate against a 1\% equivalent increment in an Earth-like planetary radius (by the presence of Venus's opaque atmosphere) in light of both spatial and temporal solar photospheric ``surface'' brightness (PSB) variations.

The solar PSB decreases radially from the heliocenter because of limb darkening. The PSB is also instantaneously non-heterogeneous on angular scales of $\sim$ 1\arcsec \ due to solar granulation, and on larger scales due to features such as sunspots.  Thus, the TSI received at ACRIMSAT (and corrected to 1AU) is expected to vary as Venus occults different portions of the photosphere during its transit due to spatial variations in PSB, separate from also expected temporal variations. 

\subsection{Photospheric Surface Brightness Variations}

We investigated the likely amplitudes of PSB variations after compensating for limb darkening that may affect ACRIM 3 measures of TSI with contemporaneous high-resolution imagery obtained with the TRACE spacecraft in its very spectrally broad WL channel ($\sim$ 0.1 -- 1.0 $\mu$m).  

We performed both temporally and spatially resolved limb-darkening corrected differential photometry of regions flanking the location of Venus as it transited the photosphere (e.g., Figure 7). With that we obtained statistical expectations of the levels of variability in TSI due to partial photospheric occultation at the angular scale of Venus (Figure 8).

{\it Temporal changes in TSI} due to Venus occultation of any fixed region of the Sun tested (e.g., denoted A-I in Fig 7) were found to be  $\pm$ 0.0018\% at the 1~$\sigma$ level (compared to a 0.0019\% expected change in signal) with inter-region variations in internal dispersions of  $\pm$ 0.00022\%. Hence, a sensitivity to the presence vs. absence of a Venus-like opaque planetary atmosphere was tested at only a 1.05~$\sigma$ level of confidence. 

{\it TSI variations due to spatial anisotropies in PSB} on Venus-size angular scales were found dispersed by $\pm$ 0.0015\% at the 1~$\sigma$ level about an expected decrement in TSI of 0.0961\% due to the presence of Venus imposed on the photosphere with compensation for limb-darkening (i.e., a 1.3 ~$\sigma$  ``detection'' of the atmosphere of Venus). 

The virtual equivalence in amplitude of the spatial and temporal variations implicates no significant systematic effects in this data set from large spatial scale PSB variations (after proper limb-darkening compensation) in excess of limiting detection sensitivities from temporal effects. We note from full-disk TRACE (and supplemental ground-based) imagery that, serendipitously, there were no large-scale photospheric structures (i.e., sunspots) which might otherwise have differentially affected the area-integrated solar brightness due to occultation by Venus. The differential effects of sunspot (or starspot)  transits, however, is anticipated and would additionally complicate transit light-curve analysis.

\section{SUMMARY}
The ACRIM 3  radiometric observation of the 5.5 hour duration ToV, with single 2.2 minute readout measures precise to one part in 10$^{-4}$  (see Fig 3) clearly demonstrate, by analogy, the ability to detect ETP stellar transits photometrically with scalably comparable instrumental sensitivities. By comparison, the goal for  {\it Kepler} differential photometry, predicated upon its expected 1 $\sigma$ noise performance estimation of $\sim$ 2 x 10$^{-5}$  for an m$_v$ = 12 solar-like star, including photon shot noise and stellar variability, will yield 4 $\sigma$ terrestrial planet detections for a single 2 -- 16 hour transit\footnote{ACRIM 3's temporal sampling of the ToV was compromised by interruptions due to Earth occultations. Photometric sampling by {\it Kepler}, which is to be placed in a solar orbit will not be so affected.} \citep{koch98}.

The amplitudes and dispersions of both the temporal and spatial PSB variations of sun-like stars would preclude discriminating with sufficient statistical significance the presence vs. absence of a Venus-like opaque planetary atmosphere for an Earth-sized transiting planet by transit observations of solar-like stars with ACRIM 3-like sampling (even if uninterrupted). Hence, following ETP detection (and orbital characterization with multi-epoch photometric observations) alternate strategies, such as spectroscopic capabilities on subsequent missions (e.g., an integral field spectrograph on TPF-C), must be considered to definitively move from detection to characterization of extrasolar terrestrial planets. 

The spatially unresolved Venus transit light curve obtained by ACRIMSAT (and a similar one obtained by SORCE/TIM) is the closest proxy to an ETP transit which exists. Given our {\it a priori} knowledge of {it our} star/planet system geometry and properties, this unique data set may be exploited  to investigate the detectability of ETP transit observation methods contemplated by future space-based terrestrial planet-finding missions. With sufficient photometric precision, proper characterization of the effects of stellar limb-darkening may yield information on the vertical structures in stellar atmospheres, and have the potential of informing on the existence of a more distended planetary atmospheres of planets transiting more quiescent stars. 

\acknowledgments
We gratefully acknowledge the cooperation and assistance of Karel Schrijver, Ted Tarbell, and other members of the TRACE and ACRIM science, planning, and operations teams  in helping us to secure the high-resolution imagery and time-resolved radiometric data used in this investigation.  GS and JMP also thank the Committee for Research and Exploration of the National Geographic Society for their support of our Venus transit studies. JMP's work is supported in part by grants from NASA's planetary and solar-terrestrial programs and the Rob Spring Fund at Williams College.

\clearpage

\begin{deluxetable}{lllll}
\tabletypesize{}
\tablewidth{0pt}
\tablehead{LIGHT CURVE SOLUTIONS}
\startdata
A) UT CONTACTS:&MEASURED&PREDICTED&\\\
\ \ \ \ CI&05h 12.95m $\pm$ 3.1m &05h 10.32m&\\\ \ \ \ CII&05h 32.42m $\pm$ 3.8m &05h 35.82m&\\\ \ \ \ CIII&10h 55.57m $\pm$ 4.4m &10h 59.25m&\\
\ \ \ \ CIV&11h 32.92m $\pm$ 5.2m&11h 29.50m&\\

B) DURATIONS:&MEASURED&PREDICTED&\% ERROR\\
\ \ \ \ Ingress&19.47m $\pm$ 4.9m &25.50m&-23.6\%\\\ \ \ \ Egress&37.35m $\pm$ 6.8m &30.25m&+23.5\%\\\ \ \ \ Transit&351.56m $\pm$ 8.4m &351.31m&+0.07\%\\

C) LIMB DARKENING&\\
\ \ \ \ {\it a}&0.85 $\pm$ 0.01\\\ \ \ \ {\it b}&0.80 $\pm$ 0.01\\

D) TRANSIT DEPTH& 0.110\% $\pm$ 0.0002\% \\
\enddata
\end{deluxetable}

\clearpage
\begin{figure}
\epsscale{1.00}
\plotone{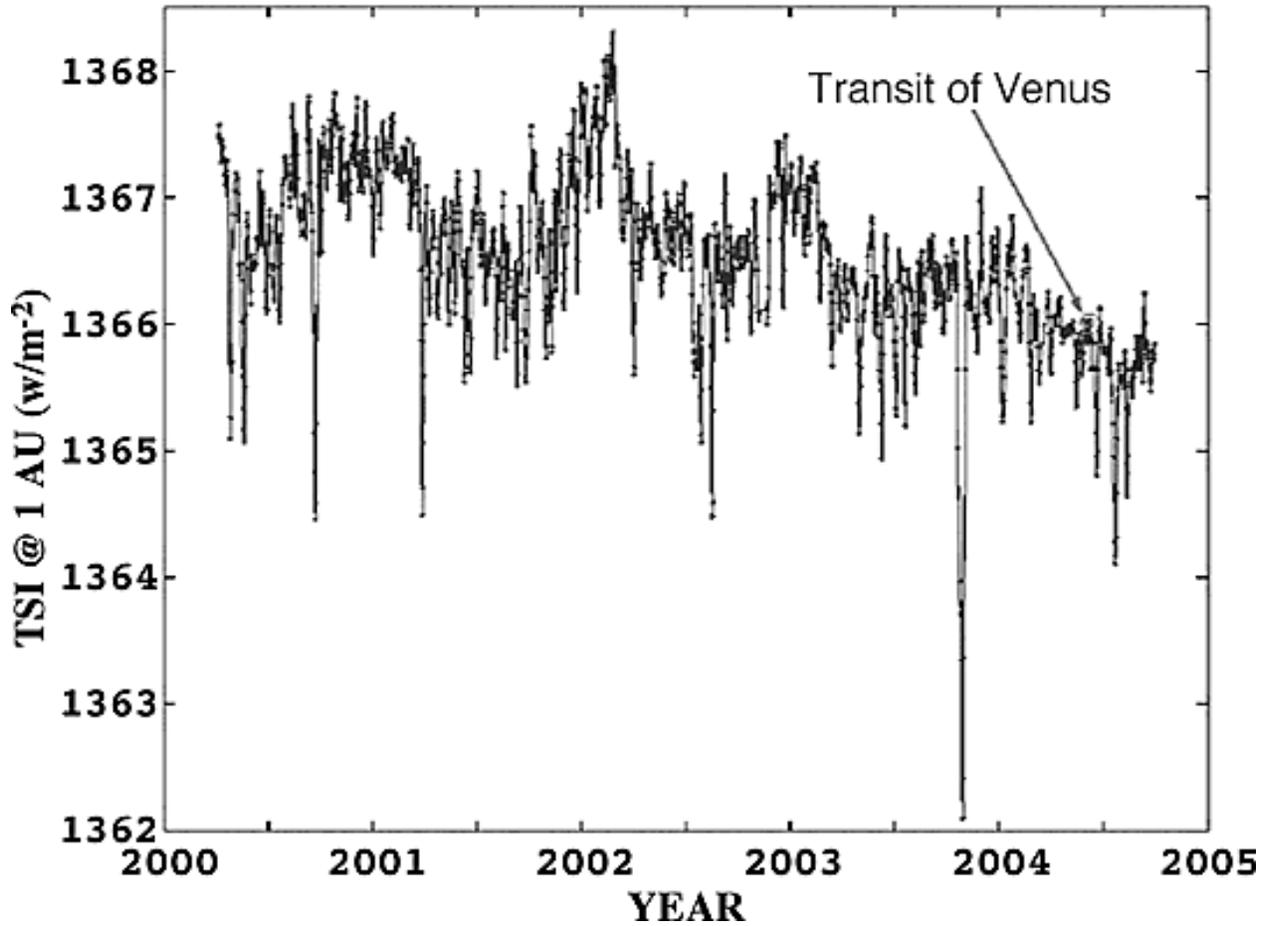} 
\caption{Daily  mean TSI (660 shutter cycle readouts per point corrected to 1 AU) from ACRIM 3 during the interval 2000.2 to 2004.7.  On 8 June 2004 (the epoch the ToV) the TSI (formerly, the ``solar constant'') was 1365.9878 W m$^{-2}$, excluding measures during the transit itself (05.0 UT to 11.5 UT).  The long-term solar variability  (on the time scales of years), of  $\sim$~0.2\%, is representative of what is expected from solar-analog targets to be observed by the {\it Kepler} mission with multiple observations of early G-type stars during the course of its 4 yr mission.\label{fig1}}
\end{figure}

\clearpage
\begin{figure}
\epsscale{1.00}
\plotone{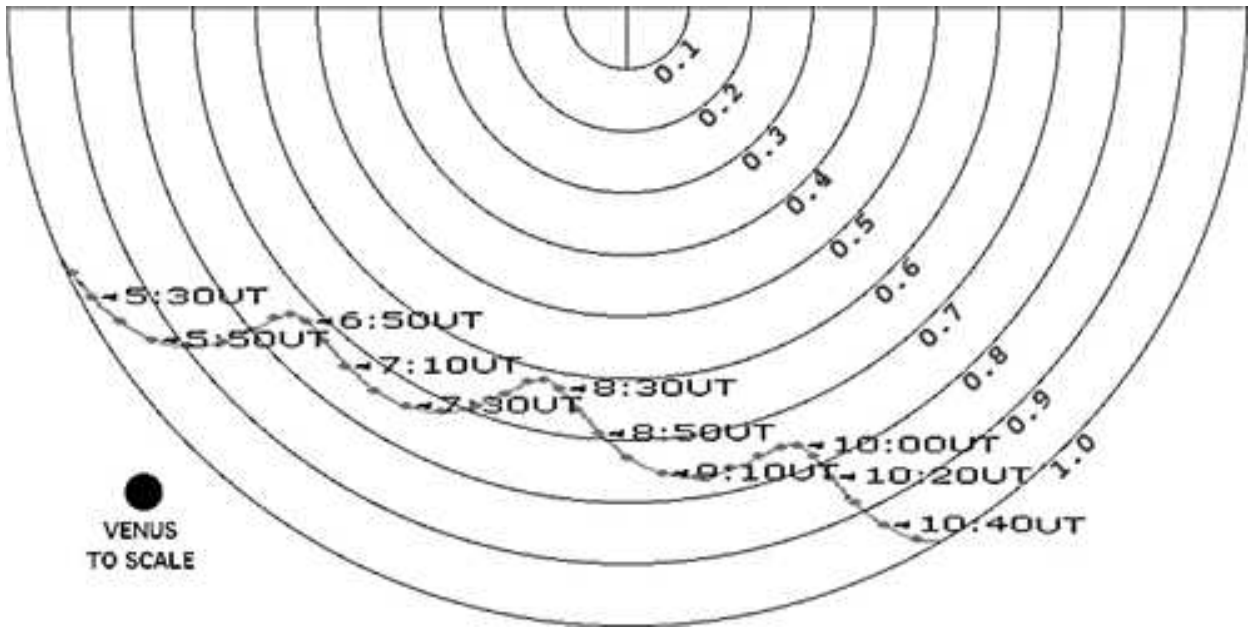} 
\caption{Apparent path of Venus crossing the face of the Sun as seen from ACRIMSAT. Points plotted in 10 m intervals illustrate the non-linear motion in the heliocentric frame.  Radial distances (in solar radii) from the heliocenter are indicated in 0.1 R$_{sun}$ increments. \label{fig2}}
\end{figure}

\clearpage
\begin{figure}
\epsscale{1.00}
\plotone{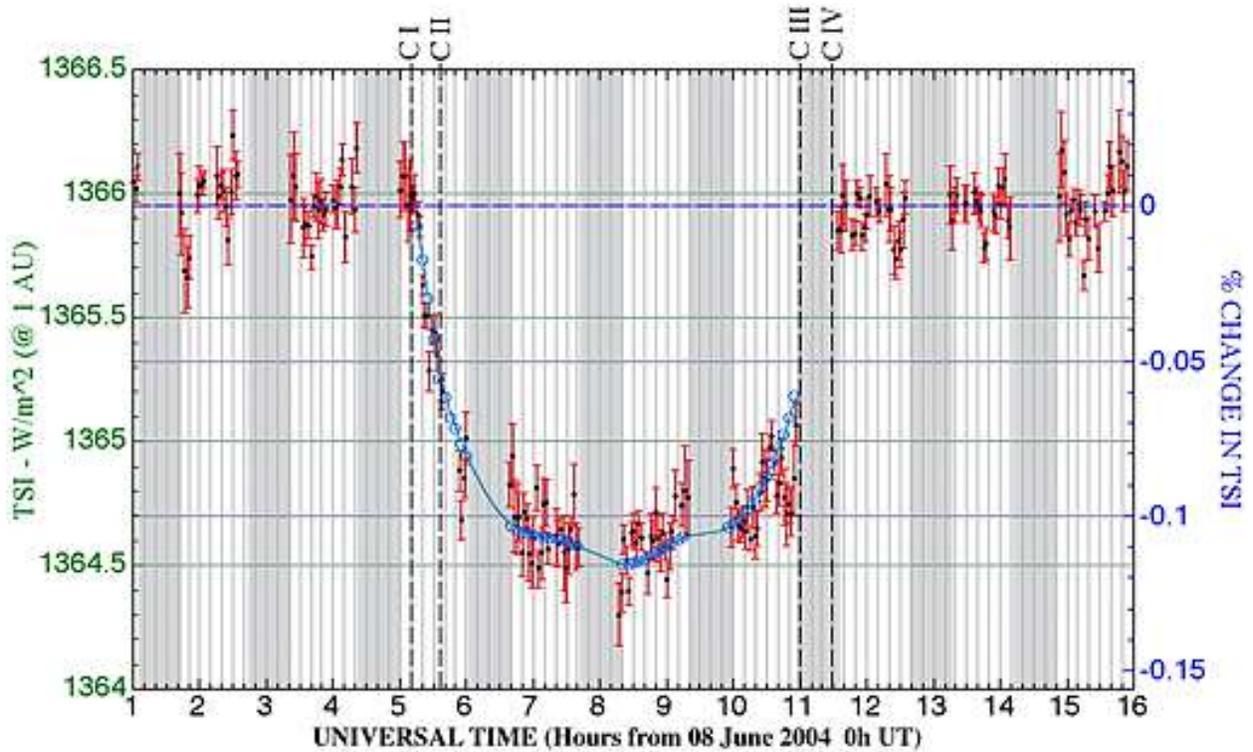} 
\caption{ACRIM 3 radiometric light curve of the 8 June 2004 ToV.  Black points, and their associated red error bars are the 131 s time-resolved ACRIM 3 TSI measures and their 1~$\sigma$ measurement uncertainties.  Blue circles are ``expected'' values from a geometrical orbit and solar limb darkening model (\S 5), during the transit at 5 m intervals while the Sun was visible to ACRIMSAT.  The Sun, as seen from ACRIMSAT, was occulted by the Earth during the times indicated by the gray bars, thus no data were available during those intervals. The vertical dashed lines labeled C1 -- CIV indicate the instants of geometrical tangency of the Cytherian and solar disks based upon the ACRIMSAT orbital ephemeris.\label{fig3}}  
\end{figure}

\clearpage
\begin{figure}
\epsscale{1.00}
\plotone{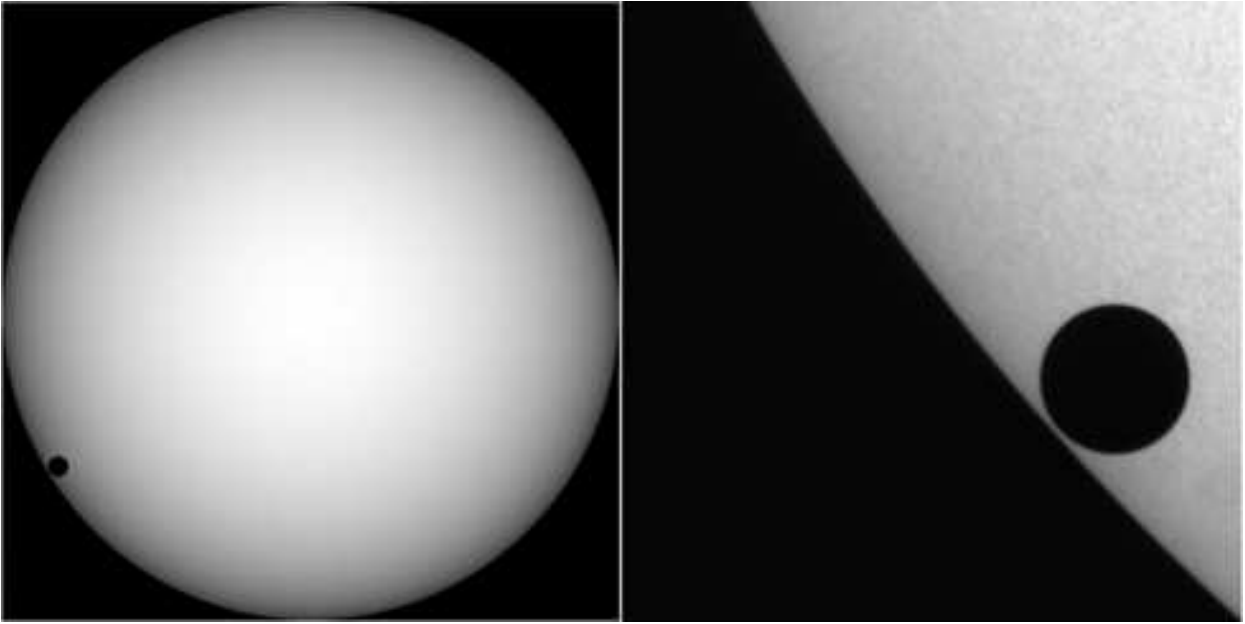}
\caption{Full-disk synthetic transit image for 05:40 UT from a ``best fit'' solution to the light curve model (left), compared to a TRACE WL image detailed in the region of Venus and the solar limb (right). \label{fig4}} 
\end{figure}

\clearpage
\begin{figure}
\epsscale{1.00}
\plotone{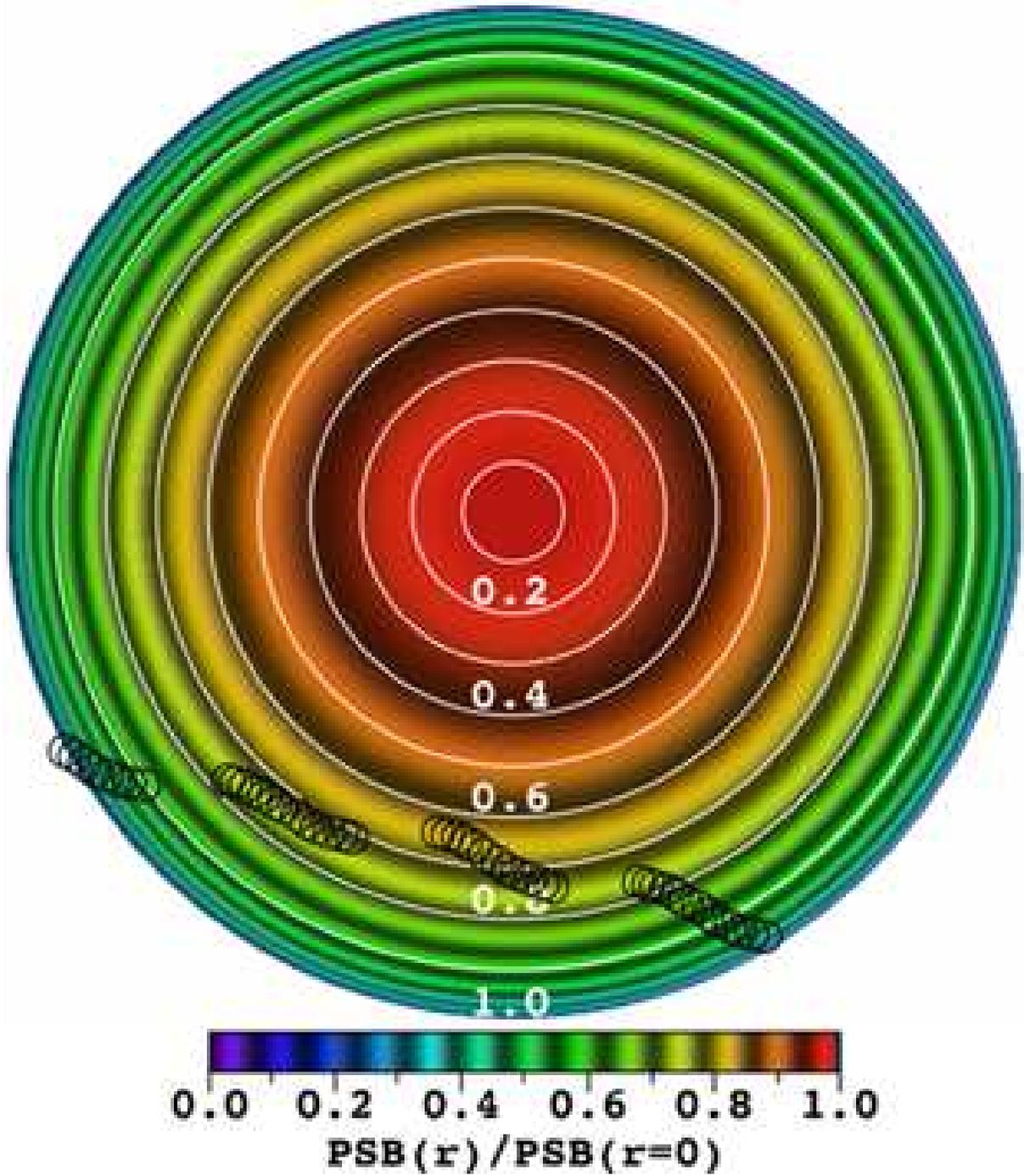} 
\caption{Center-normalized contours of photospheric surface brightness due to limb darkening characterized as described in \S 5 using {\it a} = 0.85, {\it b} = 0.80 as discussed in \S 6. White concentric circles indicate fractional distance from the heliocenter in increments of 0.1 solar radii. Small black circles indicate locations of Venus (drawn to scale) in 5 m intervals while visible to ACRIMSAT.\label{fig5}}
\end{figure}

\clearpage
\begin{figure}
\epsscale{1.00}
\plotone{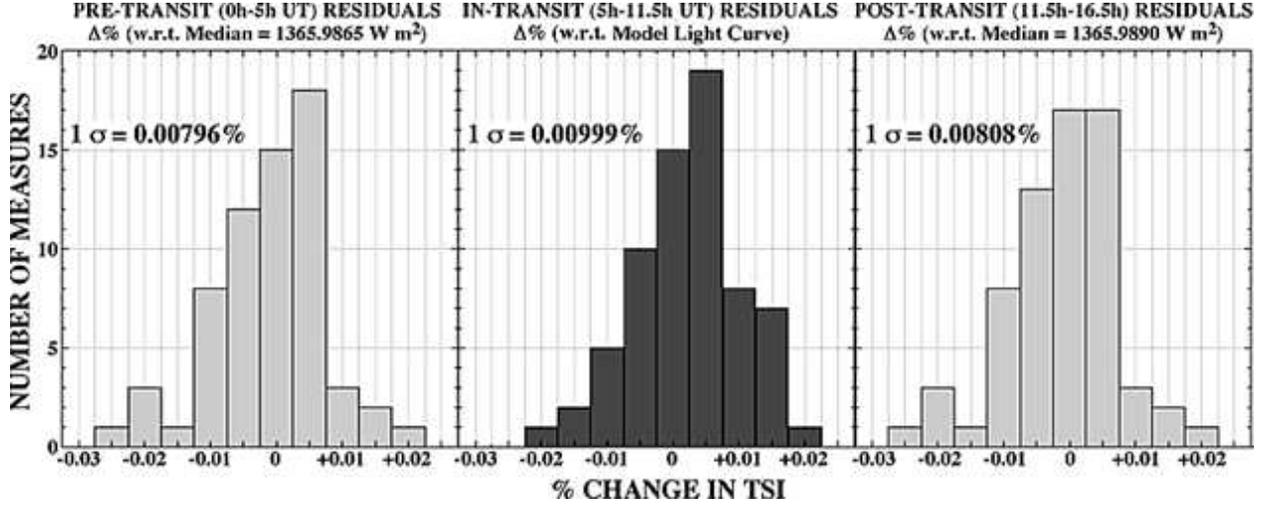} 
\caption{Left \& Right: Light curve residuals from three contiguous orbits  of pre- and post-transit data (each $\sim$ 5 hr in duration) immediately before CI and after CIV, respectively.  The median TSI measured for  those two periods differed by only 0.0002\%, i.e., is ``constant'' within the measurement errors.   The  1~$\sigma$ variation in TSI from all (unbinned) 132s ``shutter cycle readouts'' during each of these 5-hour flanking periods is $\sim$ 0.008\% of their respectively measured median TSI.  The residuals from the light-curve fit during the intervening transit (of the same duration; middle panel) are $\sim$ 0.01\% of the observed-minus-modeled TSI at the 1~$\sigma$  level (an increase of +25\% over the pre- and post-transit periods). 
The apparent increase could arise either from ({\it a}) intrinsic (temporal) solar variations during the transit, ({\it b})  systematic deficiencies in the light-curve model as discussed in \S 6.2,  or ({\it c})  shadowing, by Venus, of regions of differing photospheric surface brightness after removing the radial component due to limb darkening (\S 7).
\label{fig6}} 
\end{figure}

\clearpage
\begin{figure}
\epsscale{1.00}
\plotone{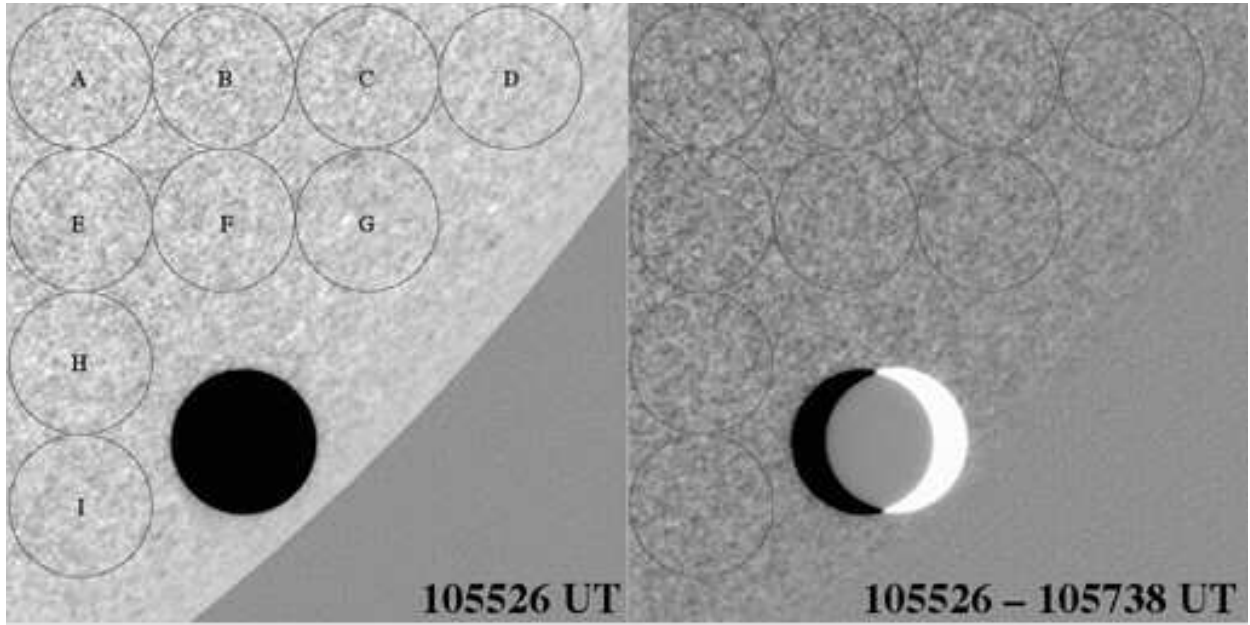} 
\caption{Left: Representative TRACE WL image, after removal of radial limb-darkening,  of Venus transiting the solar photosphere (one of 98 time-sliced images for this spacecraft pointing spanning 40 m of time). Photometric apertures (each enclosing 10,923 TRACE pixels) used to evaluate the temporal and spatial variability of the PSB on the size scale of Venus seen in projection are overlaid.  Right: Difference image  illustrating the change in PSB and the apparent movement of Venus  over the $\sim$ 131 s interval of an ACRIM 3 ``shutter cycle''  at the indicated UTs. Both images are linear displays spanning the same dynamic display range (with the sky re-normalized to mid-range of the display stretches); left: +2200 to +3200 ADU pixel$^{-1}$, right: -500 to +500 ADU pixel$^{-1}$.\label{fig7}}
\end{figure}

\clearpage
\begin{figure}
\epsscale{1.00}
\plotone{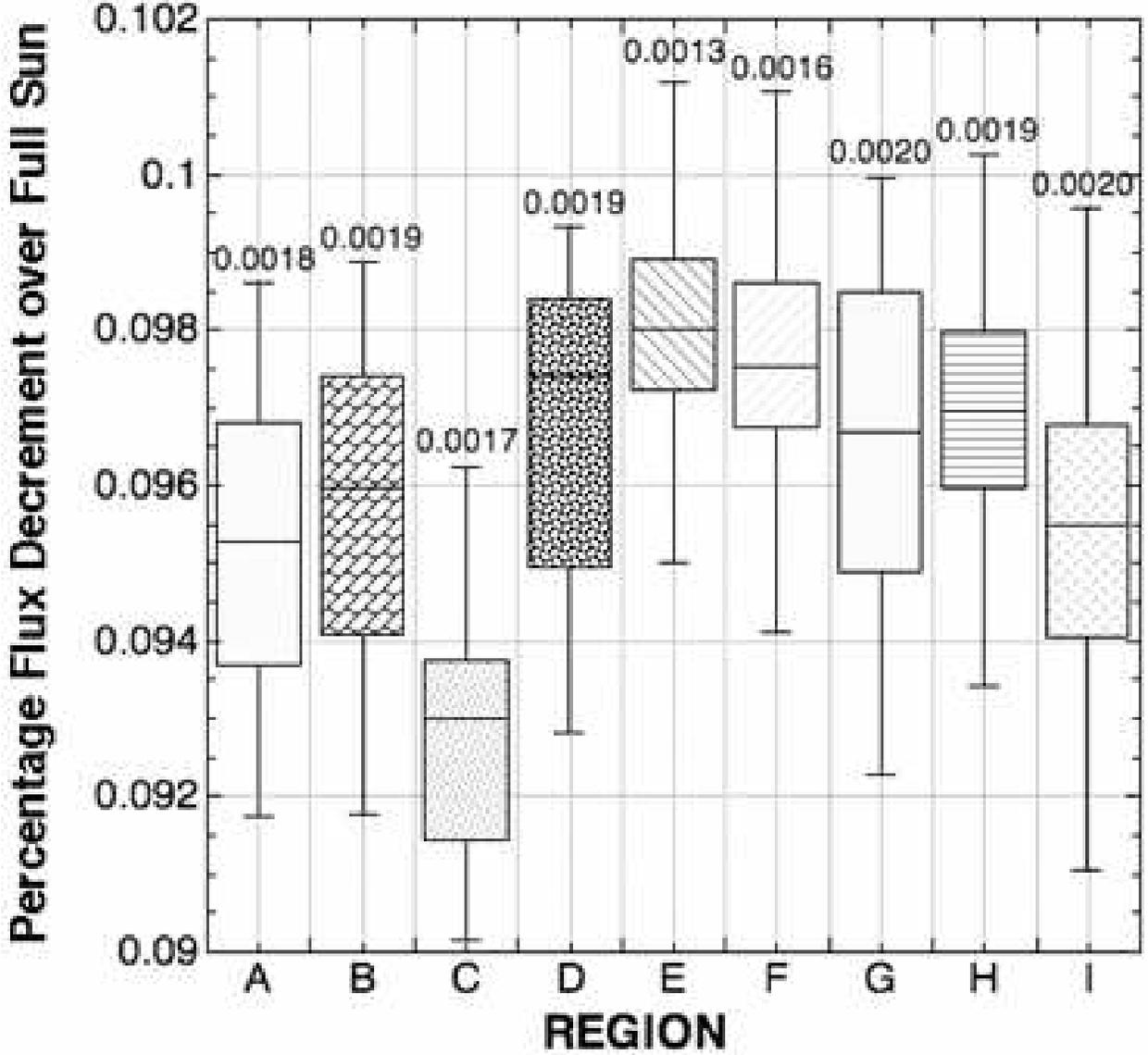} 
\caption{Variations in total solar flux density decrement ($\Delta$\% in the TRACE 0.1 to 1.0 $\mu$m passband) due to photospheric occultation by a Venus-size planet arising from temporal and spatial PSB variations. We illustrated the  variations in photometric statistics for nine representative regions on the Sun in the vacinity of Venus during the time of the transit (corresponding to the positions designated A--I in Figure 7) which would result if geometrically occulted. The height of rectangular boxes and vertical bars indicate the temporal variations in flux decrement which would result with repeated measures made (with the planetary occulter at the same position) over a 40 minute period of time.  The box half-heights 
indicate upper and lower quartiles about measured medians (black lines in boxes) of photometric measures from 98 samples.  The vertical bars indicate $\pm$ 2 $\sigma$ variations about sample means, with 1 $\sigma$ dispersions (in $\Delta$\%) annotated above each box.  Variations arising from planetary occultations of the different regions resulting from anisotropies in the solar PSB may be assessed by comparing the inter-region occultation statistics. \label{fig8}}
\end{figure}

\end{document}